\documentclass[pre,reprint,twocolumn,superscriptaddress,floatfix,aps]{revtex4-2}
\usepackage{graphicx}
\usepackage{dcolumn}
\usepackage{bm}
\usepackage{xcolor}
\usepackage{float}
\usepackage{mathrsfs}
\usepackage{soul}
\usepackage{amsmath}
\usepackage{hyperref}
\usepackage{amsfonts}
\definecolor{darkviolet}{rgb}{0.58, 0.0, 0.83}

\begin{document} 

\title{ Transitions to synchronization in adaptive multilayer networks \\ with higher-order interactions}
                    
 \author{Richita Ghosh}
  \affiliation{Department of Physics, Central University of Rajasthan, Rajasthan, Ajmer 305 817, India} 
\author{Md Sayeed Anwar}
\affiliation{Physics and Applied Mathematics Unit, Indian Statistical Institute, 203 B. T. Road, Kolkata 700108, India}
 \author{Dibakar Ghosh}
\affiliation{Physics and Applied Mathematics Unit, Indian Statistical Institute, 203 B. T. Road, Kolkata 700108, India}
\author{J{\"u}rgen Kurths}
\affiliation{Potsdam Institute for Climate Impact Research - Telegraphenberg A 31, Potsdam, 14473, Germany}
\affiliation{Humboldt University Berlin, Department of Physics, Berlin, 12489, Germany}
 \author{Manish Dev Shrimali}
 \email{shrimali@curaj.ac.in}
 \affiliation{Department of Physics, Central University of Rajasthan, Rajasthan, Ajmer 305 817, India}
 
\begin{abstract}
Real-world networks are often characterized by simultaneous interactions between multiple agents that adapt themselves due to feedback from the environment. In this article, we investigate the dynamics of an adaptive multilayer network of Kuramoto oscillators with higher-order interactions. The dynamics of the nodes within the layers are adaptively controlled through the global synchronization order parameter with the adaptations present alongside both pairwise and higher-order interactions. We first explore the dynamics with a linear form of the adaptation function and discover a tiered transition to synchronization, along with continuous and abrupt routes to synchronization. Multiple routes to synchronization are also observed due to the presence of multiple stable states. We investigate the bifurcations behind these routes and illustrate the basin of attraction to attain a deeper understanding of the multistability, that is born as a consequence of the adaptive interactions. When nonlinear adaptation is infused in the system, we observe three different kinds of tiered transition to synchronization, viz., continuous tiered, discontinuous tiered, and tiered transition with a hysteretic region.  Our study provides an overview of how inducting order parameter adaptations in higher-order multilayer networks can influence dynamics and alter the route to synchronization in dynamical systems. 
\end{abstract}

\maketitle

\section{Introduction}
Synchronization is a ubiquitous phenomenon, scattered throughout the natural and material world, such as in the flashing of fireflies \cite{buck1988synchronous}, the firing of neurons in the brain \cite{osterhage2007measuring}, power-grids \cite{rohden2012self}, etc. It is one of the most prominent collective behaviors that arise from interactions in networked dynamical systems. One of the most useful techniques in enhancing synchronization in a system is the introduction of adaptivity in complex systems. Adaptive networks are instrumental in understanding a neuronal  process called synaptic plasticity \cite{abbott2000synaptic,markram1997regulation} in which the neurons modify their connections and self-organize themselves in response to certain states such as brain damage. Adaptation-infused networks have also been studied to understand the neuronal complexity underneath Alzheimer's disease. Adaptation is also prevalent in social \cite{antoniades2015co}, epidemiological \cite{gross2006epidemic}, and geological \cite{berner2023adaptive} systems, as well as machine learning and neural networks \cite{morales2021unveiling}. The incorporation of adaptation in complex networks of phase oscillators promotes collective behavior such as clustering \cite{berner2019multiclusters},  multistability \cite{ratas2021multistability}, explosive synchronization \cite{avalos2018emergent,khanra2021explosive,xu2023dynamical}, self-organized bistability \cite{anwar2024self} and others. 
\par The route of a system attaining synchrony from a de-synchronized state may be continuous and gradual, tiered or discontinuous and abrupt. Tiered synchronization state \cite{skardal2022tiered} is referred to as the bistability of the weakly and strongly synchronized states, arising due to a pair of saddle-node bifurcations, while the bistability of incoherent and coherent states characterizes the advent of an abrupt transition. In neuronal systems, the onset of an abrupt transition may be associated with unconsciousness induced by anesthesia \cite{kim2016functional} and seizures caused by epilepsy \cite{wang2017explosive}.
\par The complexity of the neuronal network may be portrayed more realistic using multi-layer networks rather than traditional ones \cite{vaiana2020multilayer,anwar2022stability1}. Real-world networks are rarely isolated, a multi-layer network may be referred to as a ``network of networks". Multi-layer networks are used to model multi-modal data, modeling which using traditional networks can paint a simplistic and inaccurate picture of the dynamics \cite{kivela2014multilayer}. Explosive synchronization has been observed in multi-layer networks of various topologies, viz., master and slave layered networks \cite{wu2022double}, inter-pinned \cite{kachhvah2021explosive} and phase-frustrated networks along with others \cite{lotfi2018role}. The introduction of adaptation in multi-layer networks also leads to abrupt and tiered synchronization states \cite{khanra2018explosive,biswas2024effect,zhang2015explosive}.
Real-world multilayer adaptive interactions include the adaptation in a group co-ordination of orchestra \cite{boccaletti2014structure} or social networks, where the interactions at various layers lead agents to update their opinions. 
 \par However, the existing literature on multilayer adaptive networks has only considered pairwise interactions, i.e., interactions between two links or nodes. But, in the real world, interactions between multiple agents or nodes in a network occur simultaneously \cite{alvarez2021evolutionary}. For instance, in an epidemic, the behavior of agents is modeled on one layer, while the contagion process occurs on another one. As people adapt their behavior due to feedback from the situation, the dynamics of the contagion process change \cite{wan2022multilayer}. Now, it is simplistic to assume that contagion and social processes occur through pairwise interactions only as real-world interactions mostly involve multiple agents interacting with each other simultaneously \cite{battiston2021physics,majhi2022dynamics}. Such interactions are modeled mathematically as higher-order interactions using the mathematical formalism of simplices \cite{battiston2020networks}. The topology of simplices is a useful way to encode higher-order interactions in networks as it is a mathematical generalization of nodes and links. For example, a $0$-simplex is a node, $1$-simplex is a link between two nodes while a $2$-simplex is a triangle. Higher-order interactions give rise to numerous collective behaviors in systems such as enhanced synchronization \cite{zhou2006dynamical,anwar2022stability,anwar2024synchronization}, oscillation death \cite{ghosh2023first}, chimera states \cite{ kundu2022higher,ghosh2024chimeric}, turing patterns \cite{gao2023turing} and others. 
 
 \par In certain social and biological systems, individual entities adapt based on collective feedback from the system. This adaptivity can be modeled through the adaptation of the order parameter, where the system's synchrony influences the coupling strength, gradually steering the system toward eventual synchronization. An example of this is the applause of an audience, which starts with incoherent clapping and, through environmental feedback, adapts into unified applause \cite{neda2000physics}. Adaptation using the synchronization order parameters was first explored in the Kuramoto model in the context of Josephson junctions \cite{filatrella2007generalized}. This adaptation technique has been studied in many recent contributions \cite{rajwani2023tiered,biswas2024effect,zou2020dynamics} and also has been explored for multilayer networks with pairwise interactions (PI) where the authors report continuous, explosive, and tiered transitions to synchronization for different adaptation functions \cite{zou2020dynamics, biswas2024effect}. The introduction of such an order parameter adaptation in monolayer networks with higher-order interactions (HOI) has also revealed the advent of tiered synchronized states \cite{rajwani2023tiered}. 

\par In this tune, our study takes one step further and considers the impact of higher-order interactions inducted into adaptive multilayer networks in the synchronization of nonlinear oscillators. We discuss a general mathematical formalism of an adaptive multilayer network of coupled Kuramoto phase oscillators. The layers of the network are connected by means of the global order parameter, that is, each layer is adaptively controlled using the other layer's order parameter. We use the Ott-Antonsen dimensionality reduction technique \cite{ott2008low} to deduce the general reduced dimensional equations of motion for the multilayer framework with $1$ and $2$-simplex interactions and their respective adaptations. First, we consider a linear form of adaptation, i.e., layers adapt with respect to a linear functional form of order parameter. We find that applying the adaptation associated with the $2$-simplex promotes the discontinuous transition to tiered synchronization along with multiple routes to synchrony, which does not emerge in the presence of only pairwise adaptation or just higher-order interactions without adaptations. This multistability is explored by simulating the basin of attraction and checking the influence of the pairwise and higher-order coupling strengths on the system's dynamics. The route to transition is also depicted by using the continuation software MATCONT and we uncover that an extra saddle-node bifurcation occurs on the introduction of higher-order interactions apart from the subcritical pitchfork bifurcation associated with the explosive transition. Furthermore, we also consider nonlinear forms of the adaptation functions. The influence of such adaptation gives rise to both discontinuous and continuous tiered synchronized states along with multiple routes of transition to synchronization. These routes are explored along with the effect of the different exponents of nonlinear adaptations on the type of transition to synchronization. Our work sheds light on the way adaptations and higher-order interactions influence each other to produce rich dynamics in multilayer networks.

\section{Adaptive multilayer network with higher-order interactions} \label{2}
We explore the dynamics of a multilayer network with each layer composed of globally coupled Kuramoto oscillators with adaptive feedback undergoing pairwise ($1$-simplex) and three-body ($2$-simplex) interactions.
The layers are interconnected solely through adaptation functions, dependent on the global order parameter representing the coherence of each individual layer. This order parameter is defined as $r_k e^{i\phi} = \frac{\sum_{j=1}^{N} e^{i\theta_j}}{N}$. $\phi$ is the average phase of the network while $0\leq r_k \leq 1$. The phenomenon of absolute incoherence or desynchronization is denoted by $r_k \approx 0$ while $r_k \approx 1$ refers to global synchronization. 
\par The equation of motion governing the dynamics of our model system with Kuramoto oscillators on such an adaptively-coupled multilayer network of $l$ layers with $1$ and $2$-simplex interactions is represented as,
\begin{equation}
\begin{array}{l}
    \dot{\theta}_{i,l} = \omega_{i,l} + \frac{\epsilon_{1}f_{p,l}(\vec{r}(t))}{N} \sum_{j=1}^{N} sin(\theta_{j,l} - \theta_{i,l}) \\ \\ ~~~~ + \frac{\epsilon_{2}f_{h,l}(\vec{r}(t))}{N^2}\sum_{j=1}^{N} \sum_{k=1}^{N} sin(2\theta_{j,l} - \theta_{k,l} -\theta_{i,l}),
    \label{model}
\end{array}
 \end{equation}
where $i=1,2,...,N$ and $l=1,2,...,L$, with $N$ being the number of oscillators in each layer and $L$ being the number of layers. $\epsilon_{1}$ and $\epsilon_{2}$ are the pairwise and higher-order coupling strengths, respectively. $\theta_i \in [0,2\pi)$ represent the phases of the oscillators and $\omega_i \in (-\infty,\infty)$ refer to the natural frequency of the oscillator $i^{th}$, drawn from any specified distribution $g(\omega)$. $f_{p,l}(\vec{r})$ and $f_{h,l}(\vec{r})$ are the adaptation functions associated with pairwise and three-body interactions where $\vec{r}=\{r_{1,1},r_{1,2},\dots,r_{1, L}\}^T$ with $r_{1,l}=|\frac{1}{N}\sum_{j=1}^{N} e^{\mathrm{i}\theta_{j,l}}|$ being the modulus of the instantaneous complex Kuramoto order parameter of $l^{th}$ layer. 
\par For our study, we assume a two-layered network with a specific form of adaptive function given by $f_{p,1}(\vec{r})=F_p(r_{1,2})$ and $f_{p,2}(\vec{r})=F_p(r_{1,1})$, $f_{h,1}(\vec{r})=F_h(r_{1,2})$ and $f_{h,2}(\vec{r})=F_h(r_{1,1})$, where $r_{1,1}$ and $r_{1,2}$ are the global synchronization order parameters of layers $1$ and $2$. In other words, the oscillators in layer-$1$ are adaptively controlled by the order parameter of layer-$2$ only and vice-versa. We call this specific type of adaptation scheme as \textit{cross-adaptation}. A real-world example of this kind of adaptation may be found in power grids where the failure of nodes in one network contributes to failures in other networks \cite{buldyrev2010catastrophic}. In the rest of this article, we mainly focus on how this cross-adaption scheme affects the advent of the synchronization phenomenon in multilayer higher-order frameworks.

 \section{Theoretical analysis} \label{3} 
 The aim of this section is to derive a reduced (lower-dimensional) mathematical formalism that can describe the collective macroscopic dynamics exhibited by our high-dimensional model system Eq.~\eqref{model}. In order to do so, we introduce the generalized complex order parameter  $z_{m,l}$ \cite{skardal2020higher}, 
where $z_{1,l}=r_{1,l}e^{\mathrm{i}\psi_{1,l}}=\frac{1}{N}\sum_{j=1}^{N} e^{\mathrm{i}\theta_{j,l}}$ is the usual Kuramoto order parameter while $z_{2,l} = r_{2,l}e^{\mathrm{i}\psi_{2,l}}= \frac{1}{N} \sum_{j=1}^{N} e^{2 \mathrm{i} \theta_{j,l}}$ is the \textit{cluster order parameter}. Here, $r_{p,l}$, and $\psi_{p,l}$, $(p=1,2)$ are respectively the amplitude and argument of the corresponding complex order parameters. Rewriting,
 Eq.~\eqref{model} in terms of $z_{1,l}$ and $z_{2,l}$ yields 
 \begin{eqnarray}
     \dot{\theta}_{i,l} = \omega_{i,l} + \frac{1}{2i}\Big(H_l e^{-\mathrm{i}\theta_{i,l}} - H_l^* e^{\mathrm{i} \theta_{i,l}}\Big),
 \end{eqnarray}
 where $H_l=k_1z_{1,l} + k_2z_{2,l}z_{1,l}^*$, assuming $k_1 = \epsilon_1 f_{p,l}$ and $k_2 = \epsilon_2 f_{h,l}$. Here, $z_{1,l}^*$ is the complex conjugate of $z_{1,l}$.  In the continuum limit of $N\rightarrow \infty$, we represent the state of each layer by a continuous density function $f(\theta,\omega,t)$. Then, $f(\theta,\omega,t)\delta \theta \delta \omega$ denotes the density of oscillators with natural frequencies distributed between $\omega$ and $\omega+\delta \omega$ and phases between $\theta$ and $\theta+\delta \theta$. $f(\theta,\omega,t)$ satisfies the normalization condition, i.e.,
 \begin{eqnarray}
     \int_{0}^{2\pi} f(\theta,\omega,t)=1.
 \end{eqnarray}
Since the number of oscillators in each layer of the system is conserved, the density function $f$ satisfies the equation of continuity and may subsequently be written as, 
 \begin{eqnarray}
      \frac{\partial f_l}{\partial t} + \frac{\partial}{\partial \theta} (f_lv_l)=0.
      \label{eq2}
      \end{eqnarray} 
Now, $f_l$ being a $2\pi$ periodic function with respect to $\theta$, can be expanded into a Fourier series as, 
      \begin{eqnarray}
      f_l= \frac{g_{l}(\omega)}{2 \pi}\Big[1 + \sum_{n=1}^{\infty} a_{n,l} e^{in\theta_l} + \sum_{n=1}^{\infty} a_{n,l}^* e^{-in\theta_l} \Big],
      \label{eq3}
      \end{eqnarray}
where $a^*$ denotes the complex conjugate.
According to the Ott-Antonsen ansatz \cite{ott2008low}, the Fourier coefficients can be characterized as Poison kernels of the form $a_{n,l} = \alpha_l^n$, with $\alpha_{l}\ll 1$, and hence the Fourier coefficients undergo a geometrical decay for a function $\alpha$, analytic in the complex plane of $\omega$. Now, replacing the expression for $f_{l}(\theta,\omega,t)$ [Eq.~\eqref{eq3}] and $v_{l}=\dot{\theta}_{l}$ in Eq.~\eqref{eq2}, the Fourier modes collapse into a reduced $\theta$ independent single-dimensional equation for $\alpha_{l}$ which is deduced as,
\begin{eqnarray}
    \frac{\partial \alpha_l}{\partial t} = -\mathrm{i}\omega_{l} \alpha_l -\frac{1}{2}(\alpha_l^2 H_l - H_l^*),
    \label{eq7}
\end{eqnarray}
with the order parameters described as 
\begin{eqnarray}
    z_{p,l}  = \int_{-\infty}^{\infty} \alpha_l^{*p}(\omega,t)g_l(\omega)d\omega, \; p=1,2.
    \label{oa_op}
\end{eqnarray}
   \par Now, we assume a Lorentzian distribution of natural frequencies $\omega_i$ to make our model analytically tractable. The frequency distribution $g_l(\omega)$ is then given by 
\begin{eqnarray}
    g_l(\omega)&=&\frac{\Delta_{l}}{\pi}\Big[\frac{1}{\Delta^2 + (\omega_l-\omega_{0,l})^2}\Big], 
\end{eqnarray}
where for each $l$, $\omega_{0,l}$ and $\Delta_{l}$ are the mean and half-width of the frequency distribution.   

\par The order parameters given by Eq.~\eqref{oa_op} can then be derived using Cauchy's residue theorem by extending the contour to a semicircle of infinite radius in the negative half of the $\omega_{l}$ plane, which eventually provides the expressions $z_{1,l}= \alpha^*(\omega_{0,l} - \mathrm{i}\Delta_{l},t)$ and $z_{2,l}= \alpha_l^{*2}(\omega_{0,l} - \mathrm{i}\Delta_l,t)=z_{1,l}^2$.
Inserting this into Eq.~\eqref{eq7} and subsequently taking complex conjugate, we obtain the expression,
\begin{eqnarray}
    \dot{z}_{1,l} = \mathrm{i}\omega_{0,l} z_{1,l} -\Delta_l z_{1,l} + \frac{1}{2}(H_{l}^* - H_{l}z_{1,l}^2).
\end{eqnarray}
Recalling $H_{l}=k_1z_{1,l} + k_2z_{2,l}z_{1,l}^*$, the above expression can be rewritten as,
\begin{multline}
    \dot{z}_{1,l} = i\omega_{0,l} z_{1,l} -\Delta_l z_{1,l} + \frac{1}{2}[k_1z_{1,l}^* + k_2z_{2,l}^*z_{1,l} \\
    - (k_1z_{1,l} + k_2z_{2,l}z_{1,l}^*)z_{1,l}^2].
\end{multline}
Putting the relations $z_{1,l}=r_{1,l}e^{i\psi_{1,l}}$, $k_1=\epsilon_1 f_{p,l}$ and $k_2=\epsilon_2 f_{h,l}$ into the above equation, and separating the real and imaginary parts, we have the following expressions, which describes the evolution of $r_{1,l}$ and $\psi_{1,l}$ as, 
\begin{eqnarray}
    \dot{r}_{1,l} &=& - r_{1,l}\Delta_l + \frac{(1-r_{1,l}^2)r_{1,l}}{2}\Big[\epsilon_1 f_{p,l} + \epsilon_2 f_{h,l} r_{1,l}^2\Big], \\ 
     \dot{\psi}_{1,l} &=& \omega_{0,l}.
    \label{eq8}
\end{eqnarray}
Now for the ease of notation, we write $r_{1,l}$ as $r_{l}$ so the general expression for the evolution of order parameter $r_{l}$ for $l$ layers is described as,
\begin{eqnarray}
    \dot{r}_l + r_l\Delta_l = \frac{(1-r_{l}^2)r_l}{2}\Big[\epsilon_{1} f_{p,l} + \epsilon_{2} f_{h,l} r_l^2\Big].   
    \label{eq9}
\end{eqnarray}

For the sake of simplicity, we now assume that the total number of layers in the multilayer network is $L=2$. Therefore, according to Eq.~\eqref{eq9}, the evolution of the order parameters for layer-$1$ and layer-$2$ can be written explicitly as
\begin{equation}
    \begin{array}{l}
    \dot{r}_1 = - r_1\Delta_1 + \frac{(1-r_{1}^2)r_1}{2}\Big[\epsilon_{1} f_{p,1} + \epsilon_{2} f_{h,1} r_1^2\Big], \\\\
    \dot{r}_2  = - r_2\Delta_2 + \frac{(1-r_{2}^2)r_2}{2}\Big[\epsilon_{1} f_{p,2} + \epsilon_{2} f_{h,2} r_2^2\Big].
    \end{array}
    \label{op_evo_2layer}
\end{equation}
Now, we denote the steady state of Eq.~\eqref{op_evo_2layer} as $$ M_l= - r_l\Delta_l + \frac{(1-r_{l}^2)r_l}{2}\Big[\epsilon_{1} f_{p,l} + \epsilon_{2} f_{h,l} r_l^2\Big],$$ as $M_l$ for $l=1,2$ and the corresponding Jacobian to the evolution Eq.~\eqref{op_evo_2layer} is given by, 
\begin{multline}
    \frac{\partial M_l}{\partial r_j} = \Big[-\Delta_l + (\epsilon_1 f_{p,l} + \epsilon_2 f_{h,l})\frac{(1-3r_l^2)}{2}\Big]\delta_{l,j} +
    \\ \frac{(1-r_l^2)}{2}r_l\Big[\epsilon_1 \frac{\partial f_{p,l}}{\partial r_l}+ f_{h,l}\Big(\frac{\partial f_{h,l}}{\partial r_l} r_l^2 + 2r_l f_{h,l} \Big)\Big], \; l,j=1,2,
\end{multline}
where $\delta_{l,j}$ is the Kronecker delta function. 

\par Next, we proceed by applying the cross-adaptation scheme in both the pairwise and higher-order adaptation functions as
\begin{equation}
    \begin{array}{l}
          f_{p,l}=F_{p}(r_{l'}), \\
          f_{h,l}=F_{h}(r_{l'}), ~~  \;\; l,l'=1,2 ~\mbox{and}\; l \neq l' 
    \end{array}
\end{equation}
So, we set $\frac{\partial f_{p,l}}{\partial r_l}=\frac{\partial f_{h,l}}{\partial r_l}=0$. The elements of the Jacobian matrix, in this case, become as follows,
\begin{equation} 
\begin{array}{l}
    \frac{\partial M_l}{\partial r_j} \bigg|_{j=l} = -\Delta_l + \epsilon_1 f_{p,l} \frac{(1-3r_l^2)}{2} + \epsilon_2 f_{h,l} \frac{(3r_l^2-5r_l^4)}{2}, \\\\  
    \frac{\partial M_l}{\partial r_j} \bigg|_{j\neq l} = \frac{r_l \epsilon_1}{2} (1-r_{l}^2)\frac{\partial f_{p,l}}{\partial r_j}  + \frac{r_l^3 \epsilon_2}{2} (1-r_{l}^2)\frac{\partial f_{h,l}}{\partial r_j}.  
\end{array} \label{cross_jac_2layer}
\end{equation}
\par Now, from Eq.~\eqref{op_evo_2layer}, it is apparent that $(r_1,r_2)=(0,0), (r_1^*,0), \text{and} \; (0,r_2^*)$ are three trivial steady-state solutions of the aforementioned two dimensional reduced equation, where $r_l^*$ is given by, 
\begin{equation}
    \begin{array}{l}
      r_l^*=\sqrt{\frac{[\epsilon_{2}f_{h,l}(0)-\epsilon_{1}f_{p,l}(0)] \pm \sqrt{[\epsilon_{2}f_{h,l}(0)+\epsilon_{1}f_{p,l}(0)]^2-8\epsilon_{2}f_{h,l}(0)\Delta_{l}}}{2\epsilon_{2}f_{h,l}(0)}}, \\ \hfill l=1,2    
    \end{array}
\end{equation}
\par Apart from these trivial steady states, one can yield steady states of the form $(r_1^*,r_2^*) \neq (0,0)$. We numerically solve for $r_1$ and $r_2$ from the equations $(\dot{r}_{1},\dot{r}_{2})=(0,0)$, given by Eq.~\eqref{op_evo_2layer} to derive the non-trivial steady-state solutions.  
\par Now, to inspect the stability of all the steady-state solutions, we check the eigenvalues of the constructed Jacobian whose elements are given by Eq.~\eqref{cross_jac_2layer}.
\par \textbf{Stability of $(0,0)$:}
In this case, the cross-diagonal elements of the Jacobian matrix collapse to zero. Hence, the diagonal elements provide the necessary stability condition as: 
\begin{equation}
    \begin{array}{l}
\text{max} \{-\Delta_{l} + \frac{\epsilon_{1} f_{p,l}(0)}{2} \} <0, \; \text{for}\; l=1,2.  
    \end{array}
\end{equation}
Hence, it is safe to infer from the above stability conditions, that the stability of the incoherent state $(0,0)$ is independent of the higher-order interactions or its associated adaptations. 
\par \textbf{Stability of $(r_1^*,0)$:} In this case, the Jacobian matrix becomes an upper triangular matrix, and thus, again, only the diagonal elements determine stability condition.
Here, the elements of the Jacobian are given as:
\begin{equation}
 \begin{array}{l}
    \frac{\partial M_1}{\partial r_1}  = -\Delta_1 + \epsilon_1 f_{p,1}(0) \frac{(1-3r_1^{*^2})}{2} + \epsilon_2 f_{h,1}(0) \frac{(3r_1^{*^2}-5r_1^{*^4})}{2},  \\\\

    \frac{\partial M_2}{\partial r_2} = -\Delta_{2} + \frac{\epsilon_{1} f_{p,2}(r_1^*)}{2}.
\end{array}
\end{equation}
Hence the expression, 
\begin{equation}
    \begin{array}{l}
\text{max} \{\frac{\partial M_1}{\partial r_1}, \frac{\partial M_2}{\partial r_2} \} <0, 
    \end{array}
\end{equation}
gives the required stability condition.
\par \textbf{Stability of $(0,r_2^*)$:} In this case, the Jacobian matrix becomes a lower triangular matrix. Hence, the diagonal elements are able to determine the stability condition.
Here,
\begin{equation}
 \begin{array}{l}
    \frac{\partial M_1}{\partial r_1} = -\Delta_{1} + \frac{\epsilon_{1} f_{p,1}(r_2^*)}{2}, \\\\
    \frac{\partial M_2}{\partial r_2}  = -\Delta_2 + \epsilon_1 f_{p,2}(0) \frac{(1-3r_2^{*^2})}{2} + \epsilon_2 f_{h,2}(0) \frac{(3r_2^{*^2}-5r_2^{*^4})}{2}.    
\end{array}
\end{equation}
Therefore, 
\begin{equation}
    \begin{array}{l}
\text{max} \{\frac{\partial M_1}{\partial r_1}, \frac{\partial M_2}{\partial r_2} \} <0, 
    \end{array}
\end{equation}
provides the required stability condition.
\par \textbf{Stability of $(r_1^*,r_2^*)$:} Since the analytical expressions of these steady states are not tractable, we have to rely on numerical calculations to obtain the required eigenvalues of these steady states.

\section{Results}\label{4}
In order to support our analytical findings and examine the effect of the confluence of adaptivity and higher-order interactions on the emergence of synchronization in the framework of multilayer networks, we choose an order parameter adaptation function. It is expressed as
\begin{eqnarray}\label{adaptation_func}
    (Ar+B)^p,
\end{eqnarray} 
where $A,B,p \in \mathbb{R}$ and are defined as constants. When $A=1$ and $B=0$, Eq.~\eqref{adaptation_func} reduces to $r^p$, the widely discussed power-law adaptation function in networks with numerous applications \cite{filatrella2007generalized,zou2020dynamics,cai2022exact}. Now, for the bi-layered framework, we apply the cross-adaptation scheme and introduce an asymmetry in the adaptation functions of the two layers of our network, i.e. the adaptation functions associated with the pairwise interactions (referred to henceforth as $1$-simplex adaptation) in layer-$1$ expressed as $f_{p,1}=(Ar_{2}+B)^{p_1}$ and that of layer-$2$ as $f_{p,2}=(Ar_1+B)^{p_2}$. Similarly, those associated with the higher-order interactions ($2$-simplex adaptation) are referred to as $f_{h,1}=(Ar_{2}+B)^{h_1}$ and $f_{h,2}=(Ar_1+B)^{h_2}$ for layers $1$ and $2$, respectively. 
\par In our study, we explore two case studies that focus by selecting two different forms of the adaptation function: firstly, a linear functional form and secondly, a nonlinear one. In both these cases, we explore the routes of synchronization transitions by illustrating the variation of synchronization order parameters as a function of coupling strengths. To achieve this, we numerically integrated Eq.~\eqref{model} using the fourth-order Runge-Kutta method with $N=10^4$ nodes and over $20^4$ time units with time step $dt=10^{-3}$. The natural frequencies $\omega_i$ have been sampled from a Lorentzian distribution with zero mean and half-width $\Delta=1$. The global order parameter $r_l$ ($l=1,2$) has been calculated by varying the pairwise coupling strength in the adiabatic process with a step size of $0.05$ for different values of $2$-simplex coupling strength and adaptation parameters. 
 
\subsection{Linear Adaptation}\label{4.1}
We start off by considering a linear form of the adaptation function Eq.~\eqref{adaptation_func}, with the exponents $p_1=p_2=h_1=h_2=1$. Without loss of generality, the values of the constants $A$ and $B$ have been set to $2$. The constant $A$ acts as a coupling multiplier and shifts the transition point of desynchronization to synchronization, leaving the type of transition unchanged, while a non-zero $B$ results in an increase of solutions of the system and contributes to the multistability exhibited by our model Eq.~\eqref{model}. A more comprehensive depiction of the influence of $B$ on the system is given in Appendix \ref{B_effect}. 
\begin{figure}[htp]
    \centering
\includegraphics[width=0.8\columnwidth]{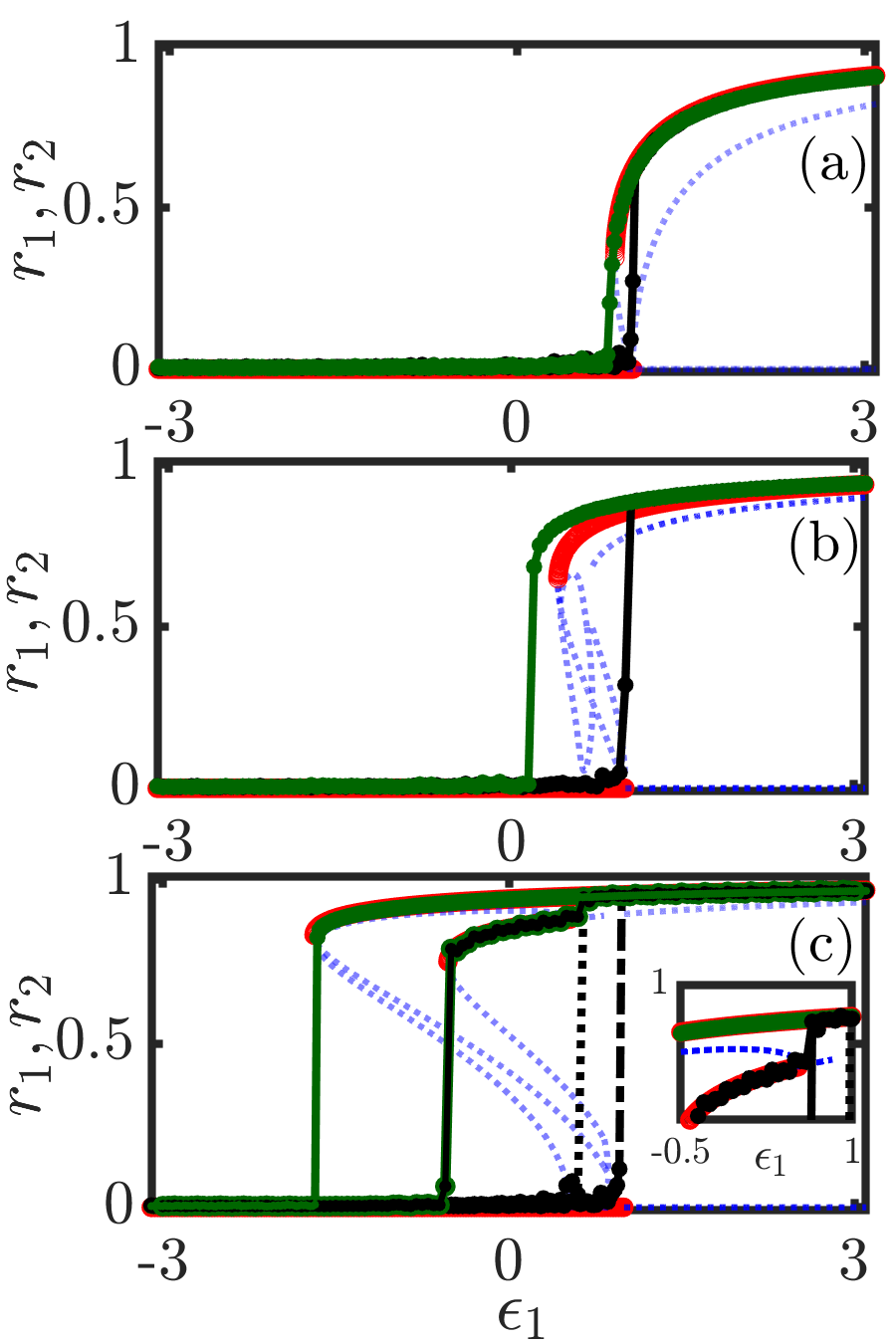}
    \caption{
    Numerical and analytical results of global order parameters $r_1$ (for layer-$1$) and $r_2$ (for layer-$2$) with varying $\epsilon_1$ with linear adaptation in three different scenarios: (a) System shows a continuous transition to sync with $1$-simplex interactions and adaptation only. (b) Explosive transition to synchronization is depicted on the application of $2$-simplex coupling without $2$-simplex adaptation, i.e., $p_1=p_2=1$, $h_1=h_2=0$ and $\epsilon_2=5.0$. (c) With both $1$ and $2$ simplices interactions and adaptions, i.e., $p_1=p_2=1,h_1=h_2=1$ and $\epsilon_2=5.0$, the system shows multiple routes to synchronization. The black and green solid lines depict the forward and backward continuations of the tiered synchronization route, the dashed lines represent an explosive route, and the dashed and dotted lines refer to another explosive route to synchronization. The dot and dashed black lines refer to one explosive pathway to synchronization, while the black dashed lines depict another. The occurrence of the tiered transition is shown in the inset. In all the panels, the green and black solid lines refer to the numerical backward and forward results, while the red solid lines and the blue dashed lines correspond to the analytical stable and unstable solutions, respectively.}
    \label{fig1}
\end{figure}
\par We investigate the dynamics of our system across three scenarios: $(1)$ with only $1$-simplex interactions and adaptation, $(2)$ with both $1$-simplex and $2$-simplex interactions but only $1$-simplex adaptation, and $(3)$ with both $1$-simplex and $2$-simplex interactions as well as adaptations. The results have been elucidated through both numerical and analytical results, which align with each other agreeably. The analytical solutions have been deduced from the eigenvalue analysis of the steady-state solutions of Eq.~\eqref{op_evo_2layer}. The stable and unstable solutions are depicted in solid red and dashed blue lines, respectively, while the results from the numerical forward and backward continuations of $r_1$ and $r_2$ are shown with green and black lines, respectively. 
\par Figure~\ref{fig1} depicts the numerical and analytical results for the evolution of the global order parameter $r_1$ ($r_2$) \footnote{Due to the symmetry in the adaptation, both the layers exhibit the same dynamics, i.e., $r_{1}$ and $r_{2}$ shows same evolution.} as a function of the pairwise coupling strength $\epsilon_1$ for the case of linear adaptation. Fig.~\ref{fig1}(a) exhibits the explosive transition with a narrow hysteric region for adaptation with only pairwise interactions. When the $2$-simplex interactions are switched on at $\epsilon_2=5.0$, but the associated adaptations are not present, i.e., $h_1=h_2=0$, we again observe an explosive transition to synchronization in Fig.~\ref{fig1}(b). But for this case, the region of hysteresis becomes wider. However, on the introduction of $2$-simplex adaptations, Fig.~\ref{fig1}(c) depicts bistability between weak and strongly synchronized states, and thus, the system shows a transition to a tiered synchronization state. Hence, the induction of the $2$-simplex adaptive functions promotes multistability in the system, i.e., the co-existence of both weak and strong synchronization states. The system jumps from desynchronization to a weakly synchronized state at around $\epsilon_1 \approx -0.5$, which eventually loses stability. This leads the system to jump to the available stable attractor, which is the strongly synchronized state. During the backward process, due to the adiabatic continuation, the system stays in the stable synchronized state and continues on it till it loses stability. The transition from incoherence to coherence during the forward process and that of coherence to incoherence in the backward process occurs at different values of $\epsilon_1$. Hence, this transition to a tiered synchronization state is accompanied by a strong hysteresis, which has not been observed during investigations into adaptive pairwise multilayer networks and monolayer adaptive networks with higher-order interactions and linear adaptation \cite{rajwani2023tiered,biswas2024effect}. Apart from this, we also observe multiple routes to synchronization. The green and black solid lines represent one transition pathway, which is the tiered transition. The dashed black lines illustrate an alternative pathway, specifically the explosive route to synchronization, while the dashed and dotted black lines exhibit another explosive route to synchronization. The system's choice of route depends on the initial conditions from which the adiabatic continuation of $\epsilon_1$ begins. Hence, the induction of even linear $2$-simplex adaptations in multilayer networks promotes the birth of tiered synchronization states as well as multiple routes to synchrony.

\begin{figure}
    \centering
    \includegraphics[width=0.8\columnwidth ]{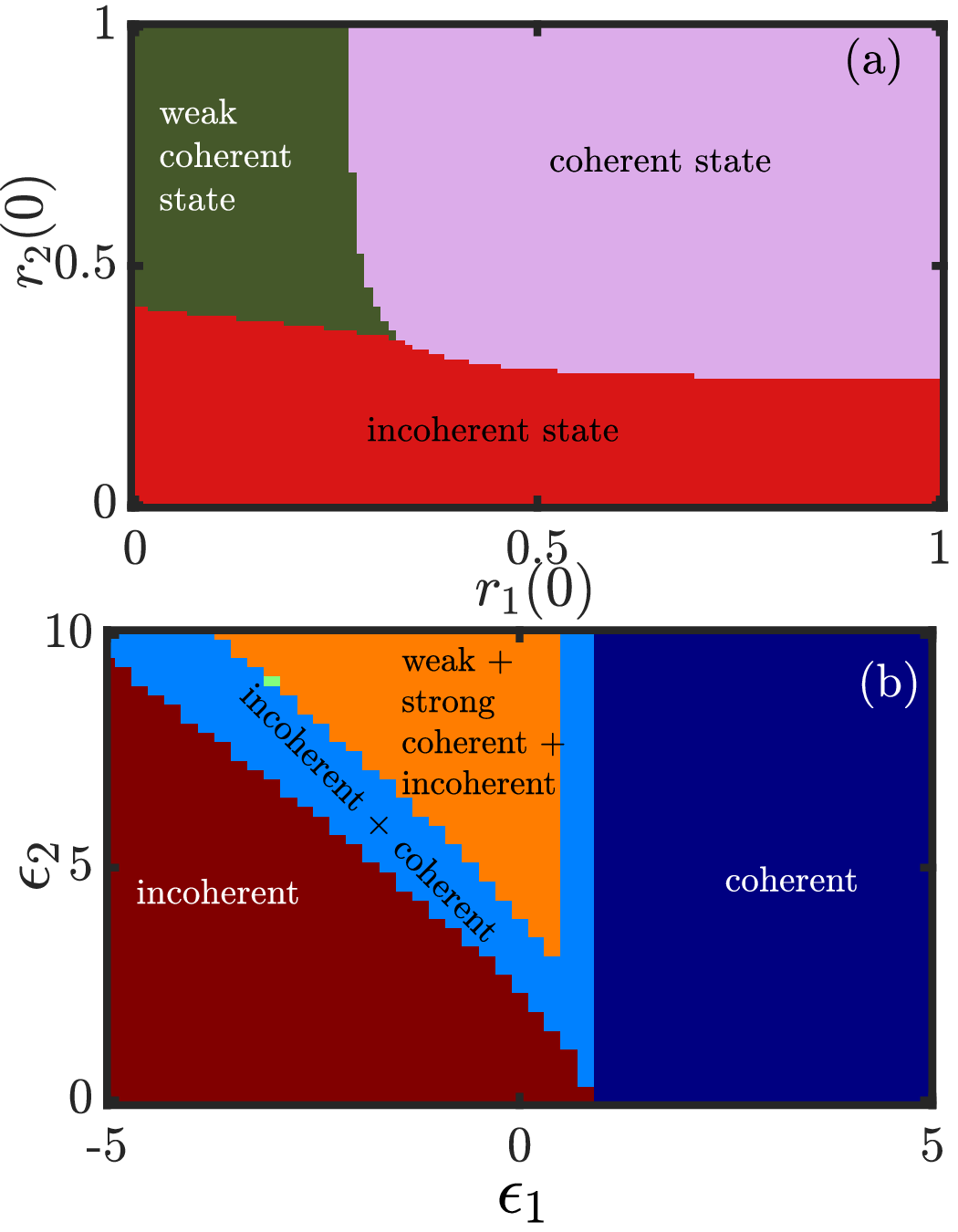}
    \caption{(a) Basin of attraction of multiple stable synchronization states observed during the tiered transition at $\epsilon_2=5$ and $\epsilon_{1}=0$. The red region depicts the initial conditions $(r_{1}(0),r_{2}(0))$ for the stable $\vec{0}$ state, the green region for the weak synchronized state, and the light purple region for the strong sync state. (b) Coherence regimes of the system as a function of $\epsilon_1$ and $\epsilon_2$. The brown regime shows the incoherent, i.e., desynchronized region, the light blue depicts the bistable region of desynchrony and synchrony, the orange region shows the multistability of desynchrony, weak synchrony, and strong synchrony while the dark blue region exhibits the entirely coherent or synchronized region. It is observable that with the increasing strength of HOI, the second-order transition to synchrony changes to the first-order, which then changes to a tiered transition subsequently.}
    \label{fig2}
\end{figure}
\par To gain a deeper understanding of the subsequent multistability arising from the tiered synchronized states, we simulate the basin of attraction (Fig.~\ref{fig2}(a)) of the stable attractors of the system with linear adaptation at $\epsilon_2=5.0$ and $\epsilon_1=0.0$ for the reduced equations of motion of $r_1$ and $r_2$, i.e., Eq.~\eqref{op_evo_2layer}. The values of $r_{1}$ have been plotted for the basin. In Fig.~\ref{fig2}, $r_1(0)$ and $r_2(0)$ refer to the initial conditions of the first and second layers respectively. The red regime shows the initial conditions for which the system stays in the stable $\vec{0}$ or the incoherent state. The dark green depicts the initials for the weak coherent state, while the light purple color exhibits that of the strong sync state. It is important to note that if we plot $r_2$ instead of $r_{1}$, then a mirror symmetry of the basin along the left diagonal will be obtained (not shown in this figure). In other words, the regions of the weak coherent state that are shown in the plot (Fig.~\ref{fig2}(a)) correspond to the trivial steady states of the form $(a,0)$, while the region of weak coherent state associated with $r_{2}$ will correspond to the steady states of the form $(0, a)$. From the basin, we can further deduce the initial conditions required for the multiple stable states. For example, it can be observed that the weak synchronized state can be achieved by setting the initial conditions of the oscillators in one layer to be in incoherence ($r_{1}(0) \approx 0$) and that in the other layer to be in coherence ($r_{2}(0) \approx 1$).  

\par Next, we explore the two-dimensional bifurcation diagram of the $1$-simplex coupling strength $\epsilon_1$ and $2$-simplex coupling $\epsilon_2$ through Fig.~\ref{fig2}(b). Varying $\epsilon_1$ and $\epsilon_2$, we obtain the different regimes of synchronization that the system exhibits. Without any HOI, the system makes a smooth and continuous transition from the incoherent state (brown region) to the coherent state (dark blue region). However, the introduction of even a weak $\epsilon_2$ brings about the appearance of a bistable region (in blue), which gets wider as $\epsilon_2$ is increased. Thus, even a HOI coupling of very low strength encourages the system to undergo a first-order transition to synchrony. When the strength of $\epsilon_2$ is increased even more, we find the appearance of an orange regime depicting the multistability of incoherent, weakly coherent, and strong coherent states. Hence, the increasing strength of HOI changes the transition type from second-order to first-order and subsequently to tiered one. 
\begin{figure}[htp]
    \centering
    \includegraphics[width=0.5\textwidth ]{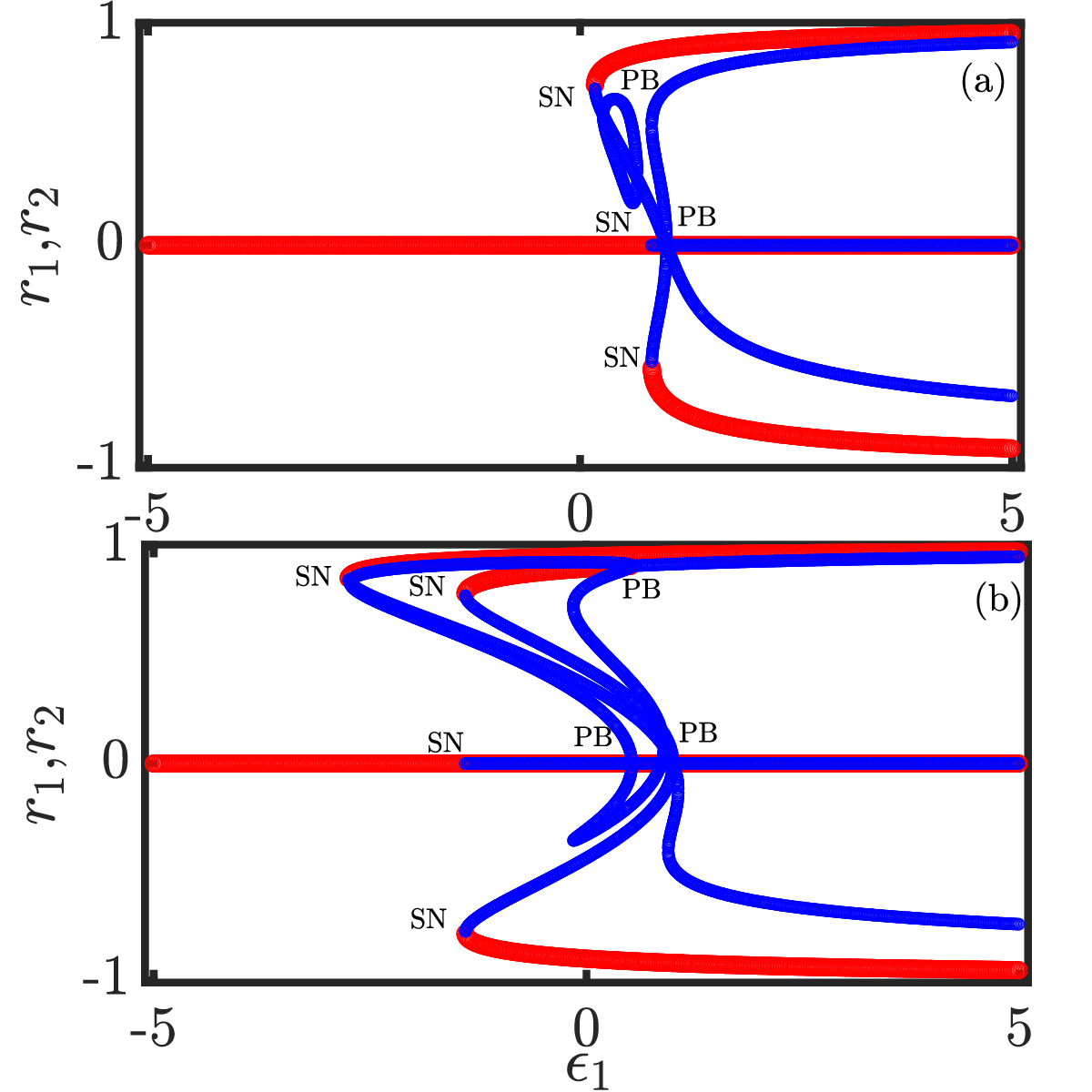}
    \caption{Routes to synchronization of the adaptive multilayer network of Kuramoto oscillators have been depicted through bifurcation diagrams. Analytical curves of order parameters $r_1$ and $r_2$ are drawn in red for the stable branches and in blue for the unstable branches. Here, $SN$ and $PB$ stand for saddle-node and pitchfork bifurcation, respectively. (a) The system shows an explosive transition to synchronization through the subcritical pitchfork bifurcation with both pairwise and non-pairwise interactions but with no adaptation associated with HOI. Parameters: $p_1=p_2=1$ and $h_1=h_2=0$, $\epsilon_2=5$. (b) Tiered synchronization is depicted due to two pitchfork and saddle-node bifurcations in a network of linear adaptation associated with both PI and HOI. Parameters: $p_1=p_2=1$ and $h_1=h_2=1$, $\epsilon_2=5$.}
    \label{fig3}
\end{figure}
\par  We also investigate the route to the various synchronization states of our system with linear adaptation through bifurcation analysis. To explore the various bifurcations occurring in the system, we have used the continuation software MATCONT \cite{dhooge2003matcont}. In order to present a complete picture of the bifurcation diagram, we have included the negative and inadmissible values of $\vec{r}$ also.  We first switch off the $2$-simplex adaptation, i.e., $h_1=h_1=0$ but the HOI coupling strength $\epsilon_2=5$ is present. We observe a discontinuous first-order transition from desynchronization to synchronization through a subcritical pitchfork bifurcation (PB) and subsequent saddle-node bifurcations (SN) in Fig.~\ref{fig3}(a). We also note the presence of a reverse subcritical pitchfork bifurcation at the same point that leads to the creation of the two unstable branches. Figure~\ref{fig3}(b) exhibits the bifurcations leading to the creation of the tiered synchronization state. With the introduction of the $2$-simplex adaptation, an extra saddle-node bifurcation occurs, and the weak synchronization state is born as a consequence. On further increase of $\epsilon_1$, this weak state of synchrony suffers a reverse subcritical pitchfork bifurcation, and the stability of the branch changes from stable to unstable through a second subcritical pitchfork bifurcation, leaving only the strong synchronization state to persist for higher values of $\epsilon_1$. These additional saddle-node and subcritical pitchfork bifurcations lead to the appearance of the tiered transition to synchronization and are solely a consequence of the induction of the $2$-simplex adaptations. In order to present a succinct picture of the bifurcations occurring in the system, we have also included the negative and non-admissible solutions of $\vec{r}$ in our figure. 

\begin{figure*}[htp]
    \centering
    \includegraphics[width=\linewidth ]{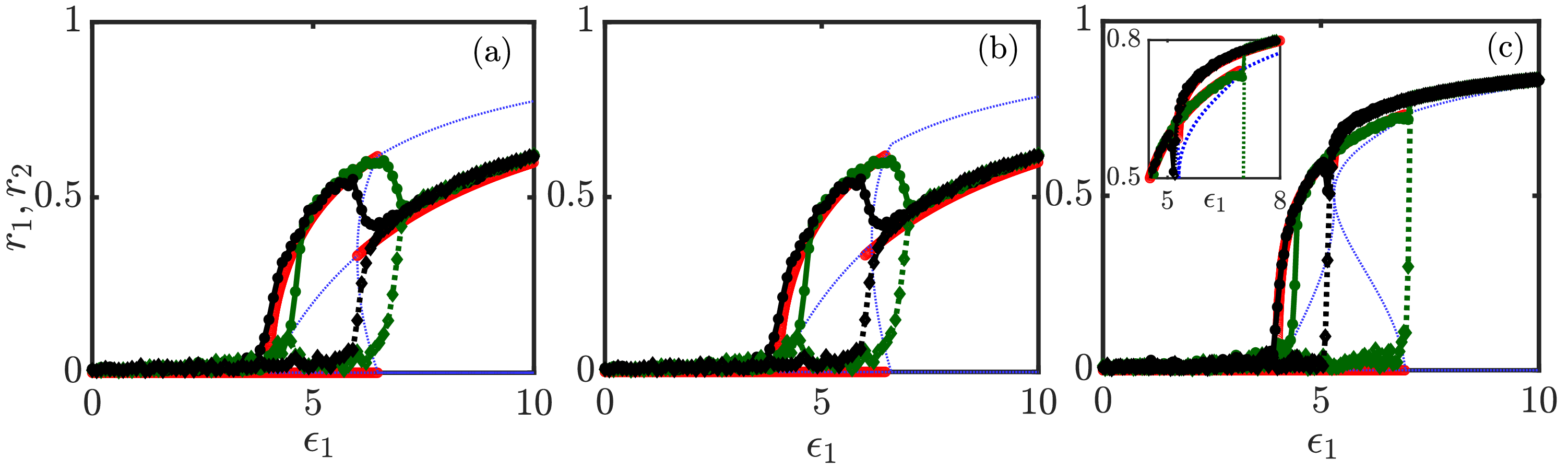}
    \caption{Numerical and analytical results for the model system with nonlinear adaptation are presented. The green and black solid lines represent the forward and backward continuations for one set of initial conditions, while the dashed green and black lines represent the same for another set of initial conditions. The solid red and dashed blue lines refer to the stable and unstable analytical solutions, respectively: (a) for $1$-simplex interactions and adaptation with $p_1=p_2=-1$ and $h_1=h_2=0$ at $\epsilon_2=0$. The system transitions to sync and, subsequently drops to a weaker sync state with increasing $\epsilon_1$. (b) Transition to synchronization for the system with $1$-simplex interaction and adaptation with only $2$-simplex interaction with $p_1=p_2=-1$ and $h_1=h_2=0$ at $\epsilon_2=0.4$. (c) Transition to tiered synchronized state for $1$ and $2$ simplices interactions and adaptations with $p_1=p_2=-1$ and $h_1=h_2=2$, at $\epsilon_2=0.4$. The zoomed-in picture of the tiered transition is included in the inset. }
    \label{fig4}
\end{figure*}
\subsection{Nonlinear Adaptation}\label{4.2}

Now, we explore how the introduction of nonlinearity in the adaptation affects the synchronization transitions in the multilayer framework Eq.~\eqref{model}, i.e., we choose the exponents $p_1,p_2,h_1$, and $h_2$ such that the adaptation function Eq.~\eqref{adaptation_func} is nonlinear.
\par i) At first, we choose a symmetric form of adaptation in the layers, i.e., the exponents of $1$- and $2$-simplex adaptations are equal for both the layers. We set the parameters $p_1=p_2=-1$ and $h_1=h_2=2$.  Figure~\ref{fig4}(a) depicts the dynamics of the system with $1$-simplex interaction and adaptation with a changing $\epsilon_1$. We observe that the system undergoes a ``reverse tiered synchronization" where the stable synchronized branch drops down from a strong to a weaker one when the adiabatic process is started from one set of initial conditions. We also observe an alternative route to synchronization which is depicted by the dashed green and black lines showing an explosive transition for another set of initial conditions. Hence, the symmetric form of adaptation in multilayer networks promotes multiple routes to synchronization.  When adding the $2$-simplex interaction $\epsilon_2\approx0.4$, the dynamics of the system do not change and remain identical to that of the system with only PI (Figure~\ref{fig4}(b)). However, when $2$-simplex adaptations are also added along with the interactions, we observe that the system undergoing a continuous tiered transition to synchronization, i.e., the system exhibits a continuous transition to synchrony and then subsequently, as this synchronized branch becomes unstable with increasing $\epsilon_1$, it jumps up to meet a stronger branch of synchronization. We also find multiple transitions to synchronization since, for certain initial conditions, the system undergoes a tiered transition, and for others, it exhibits an explosive transition to synchrony. To gain a clearer understanding of the multistability associated with the tiered state, we have simulated its basin of attraction, as detailed in Appendix \ref{basin_nonlinear}.
\begin{figure*}
    \centering
    \includegraphics[width=\linewidth]{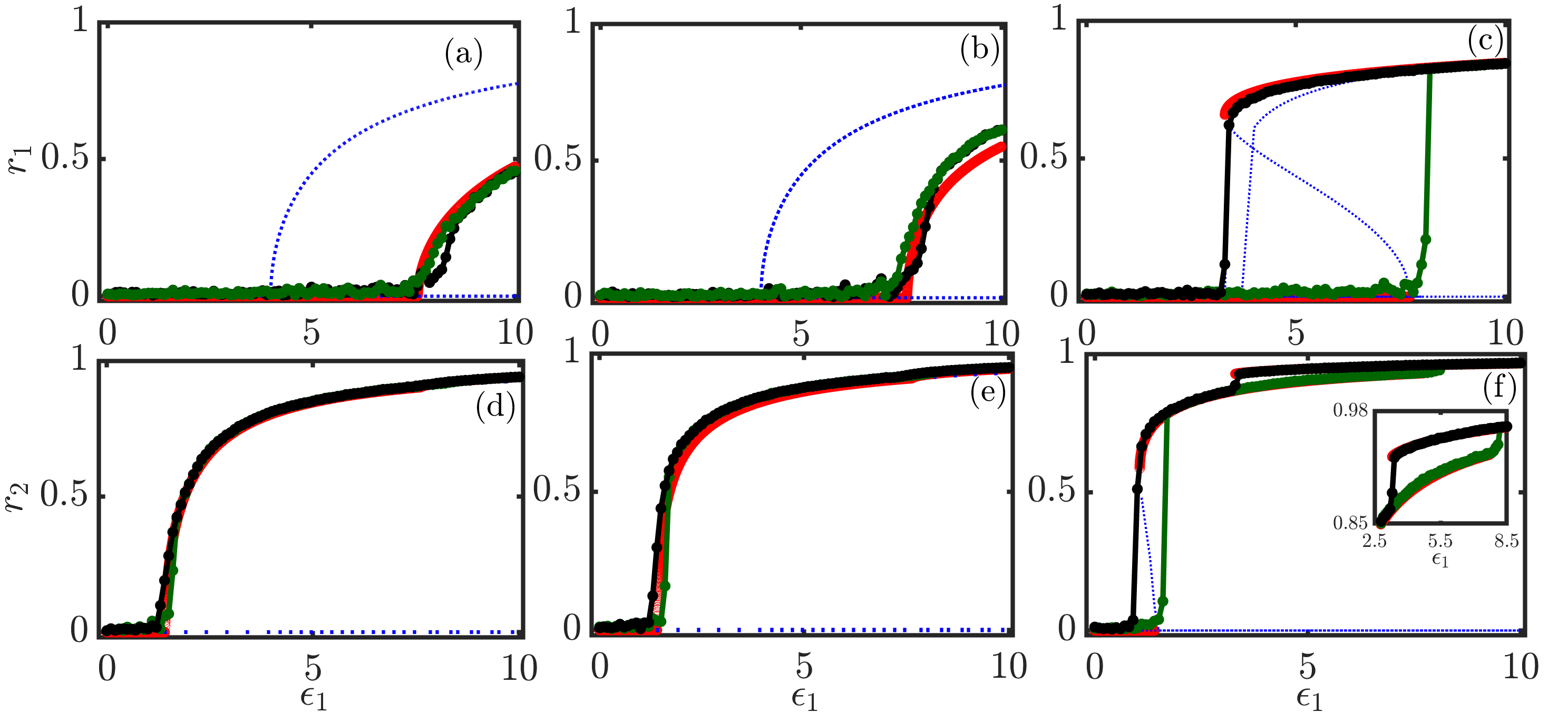}
    \caption{Numerical and analytical results for the model system with nonlinear adaptation. Left panel: (a,d) For $1$-simplex interactions and adaptations $p_1=-1,p_2=1$, a second-order transition to synchrony is observed for both layers $1$ and $2$. Middle panel: (b,e) For $1$-simplex interactions and adaptations with only $2$-simplex interaction at $\epsilon_2=1.6$, continuous transitions to synchronization are exhibited by layer $1$ and layer-$2$. Right panel: (c) Explosive transition to synchronization is observed for $p_1=-1,p_2=1,h_1=1.0,h_2=1.5$ at $\epsilon_2=1.6$. (f) Discontinuous tiered synchronization is depicted for the layer-$2$ with a discontinuous transition from the stable incoherent state to a weaker branch of synchronization and a subsequent discontinuous transition from this weak branch to a stronger one. The tiered transition portion is displayed in the inset.}
    \label{fig5}
\end{figure*}
\par ii) Next, we examine the dynamics of the system with asymmetry in the choices of adaptations for both layers $1$ and $2$. We choose the adaptation exponents as $p_1=-1,p_2=1,h_1=1.0,$ and $h_2=1.5$. In Fig.~\ref{fig5}, the entire upper and below panel shows the dynamics of layer $1$ and layer-$2$, respectively, as a function of $\epsilon_1$ under various conditions. Figures~\ref{fig5}(a),~\ref{fig5}(d) depict continuous transitions to synchronization when the system is equipped with $1$-simplex adaptations only. When the $2$-simplex interaction is applied with $\epsilon_2=1.6$, a similar second-order transition is observed on inspecting global order parameter $\vec{r}$ varying with $\epsilon_1$ (Figs.~\ref{fig5}(b),~\ref{fig5}(e)). With the application of $2$-simplex adaptations, Fig.~\ref{fig5}(c) depicts an explosive transition to synchrony in layer-$1$ while Fig.~\ref{fig5}(e) shows a discontinuous tiered transition to synchrony occurring in layer-$2$ with the explosive transition to a weaker branch of synchronization takes place at around $\epsilon_1 \approx 2$. Subsequently, another discontinuous transition occurs to a stronger branch of synchrony from the weaker one. We also investigate the routes to synchrony through bifurcation analysis to attain a deeper understanding of the underlying dynamics of the system. Figure~\ref{fig6} illustrates the changes in the bifurcations in the systems with the application of the $2$-simplex adaptations. Figures~\ref{fig6} (a),~\ref{fig6}(d) note the supercritical pitchfork bifurcations occurring in layers $1$ and $2$ when only the $1$-simplex adaptations and interactions are present. The application of a $2$-simplex interaction with no $2$-simplex adaptation, i.e., $h_1=h_2=0$, does not alter the road to synchronization, and the bifurcations remain identical (Figs.~\ref{fig6}(b),~\ref{fig6}(e)). On the other hand, Fig.~\ref{fig6}(c) shows that the addition of $2$-simplex adaptations alters the super-critical pitchfork bifurcation to a subcritical one in layer-$1$. The same occurs in layer-$2$ as illustrated in Fig.~\ref{fig6}(e), along with the occurrence of an extra saddle-node bifurcation, which leads to the creation of the tiered synchronization state. It is to be noted that the non-admissible negative solutions and some of the positive solutions corresponding to these negative solutions are also depicted in Fig.~\ref{fig6} to paint a clearer picture of the bifurcations.
\begin{figure*}[htp]
    \centering
    \includegraphics[width=\linewidth ]{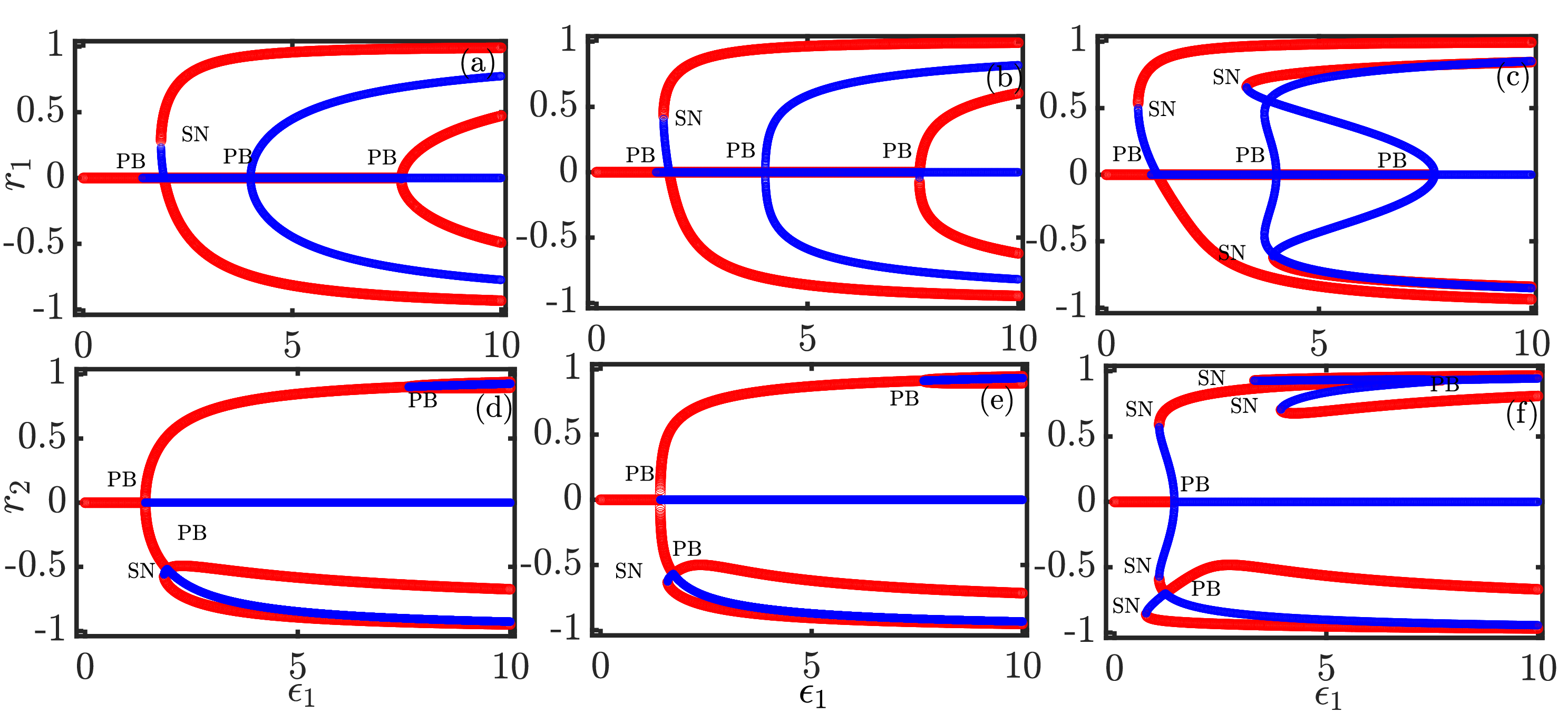}
    \caption{Bifurcation diagram for the system with nonlinear adaptation: The bifurcation diagram is depicted through stable and unstable analytical lines in red and blue solid lines, respectively. $SN$ and $PB$ represent saddle-node and pitchfork bifurcations, respectively. Left panel: (a,d) represent the bifurcation diagram for $p_1=-1,p_2=1$ and with $\epsilon_2=0$. Supercritical pitchfork bifurcation occurs at the point of transition from desynchrony to synchrony in layers $1$ and $2$. Middle panel: (b,e) represents the bifurcation for $1$-simplex adaptation and interaction with $\epsilon_2=1.6$, and the dynamics are identical to (a) and (d). Right panel: (c),(f) depict the bifurcations for $p_1=-1,p_2=1,h_1=1.0,h_2=1.5$ and $\epsilon_2=1.6$. (c) Subcritical pitchfork and saddle-node bifurcations for the explosive transitions from desynchronization to synchronization. (f) Two subcritical pitchfork and subsequent saddle-node bifurcations occur for the discontinuous tiered transitions to synchronization. }
    \label{fig6}
\end{figure*}
 
\begin{figure}[htp]
    \centering  \includegraphics[width=\columnwidth]{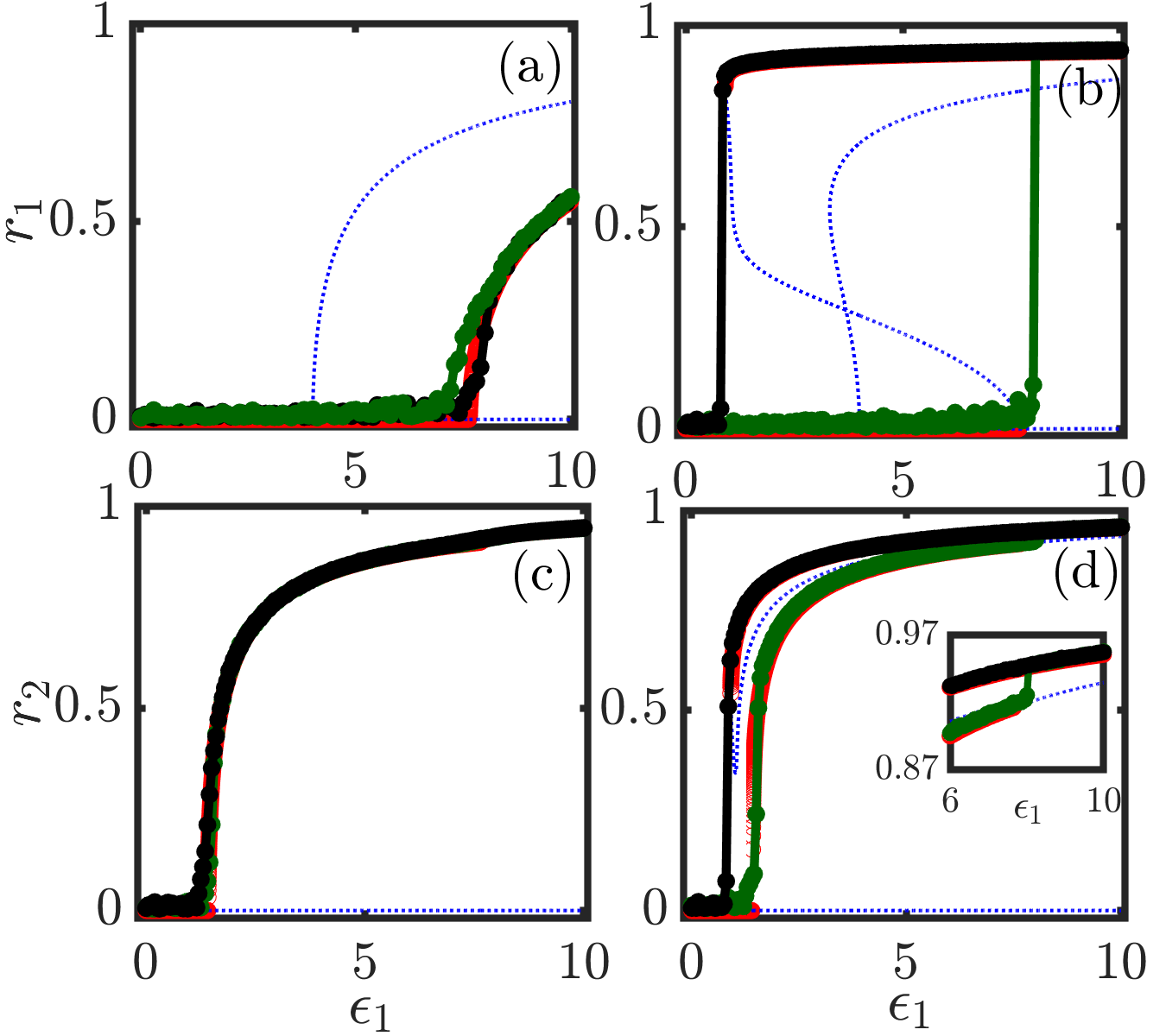}
    \caption{Numerical and analytical results for the model system with nonlinear adaptation. Left panel: (a,c) For $1$-simplex interactions and adaptations $p_1=-1,p_2=1$ with only $2$-simplex interaction at $\epsilon_2=1.0$, second-order transition to synchronization is observed for both layers $1$ and $2$. Right panel: (b) For both $1$ and $2$-simplex interaction and adaptations $p_1=-1,p_2=1$ and $h_1=1.0,h_2=1.5$ at $\epsilon_2=1.0$, explosive transition to synchronization is exhibited by layer $1$. (d) Tiered synchronization with a hysteretic region is depicted for layer-$2$ with a discontinuous transition from a weaker branch of synchronization to a stronger one. The tiered transition is shown in the inset. }
    \label{fig7}
\end{figure}
\par We further explore the asymmetric adaptation scenario by changing the higher-order coupling strength $\epsilon_{2}$, while keeping fixed the exponents of the adaptation function the same as in Fig.~\ref{fig6}, i.e., $p_1=-1$, $p_2=1$, $h_1=1.0$, and $h_2=1.5$. Figure~\ref{fig7} shows the corresponding results for $\epsilon_2=1.0$. In Figs.~\ref{fig7}(a),~\ref{fig7}(c), the order parameters $r_1$ and $r_2$ exhibit a second-order transition to synchronization in first and second layers with $p_1=-1,p_2=1,h_1=0$, and $h_2=0$. Figure~\ref{fig7}(b) displays that the order parameter $r_{1}$ goes through a system and undergoes an explosive transition to synchronization with $p_1=-1,p_2=1,h_1=1.0,h_2=1.5$ and $\epsilon_2=1.0$. While inspecting $r_2$, we find a continuous tiered transition to synchrony with a hysteresis region (Fig.~\ref{fig7}(d)). In this layer, the system undergoes a continuous transition to synchronization, which subsequently becomes unstable, and the system jumps to a stronger branch of synchrony. In the backward process, the system comes down to desynchrony following a different branch of synchronization than the one followed for the forward process. Hence, a hysteresis area is observed due to the occurrence of the tiered synchronization. 
\section{Conclusion}
Adaptive networks have been a focus issue in the study of complex networks in recent times. While the dynamics of adaptive multilayered networks with pairwise interactions have been explored previously in some articles \cite{zhang2015explosive,biswas2024effect}, non-pairwise interactions in such networks generate much richer dynamics. 
\par Our study considers both $1$ and $2$-simplex interactions and their corresponding adaptation functions in a multilayered (two-layered) system of Kuramoto oscillators. The oscillators in the layers are adaptively controlled through the global synchronization order parameter of other layers. Firstly, we have considered linear forms of adaptations. The application of $2$-simplex interactions and adaptations gives rise to tiered synchronization states due to multistability and multiple discontinuous transitions, as opposed to the explosive synchronization seen with only higher-order interactions. We have explored the multistability of the system by inspecting the basin of attraction of the stable attractors and also noted the influence of $1$ and $2$-simplex interactions on the type of synchronization states the system exhibits. The routes to such synchronized states have also been investigated, and the tiered states of synchrony have been observed to be born due to a pair of saddle-node bifurcations. 
\par Furthermore, we have also explored the system by considering a nonlinear form of adaptation functions. Three sets of adaptation exponents have been chosen, and in every case, we have uncovered a unique type of tiered synchronization state, apart from explosive transitions. For the first case, we have observed the coexistence of a continuous transition to a tiered synchronization state and an explosive transition as well. The second case exhibits a discontinuous transition to the tiered synchronization state in one layer and explosive synchronization in the other on the application of asymmetric $2$-simplex adaptations in both layers. The bifurcations in this case are also explored. As with the linear adaptation case, we have found a subcritical pitchfork bifurcation leading to the explosive transition to synchronization and a couple of saddle-node bifurcations creating the tiered synchronization states. The third case depicts a continuous transition to a tiered state with a hysteretic region, i.e., there is bistability of weakly synchronized branches in this layer. We also explore the combined influence of the higher-order interaction strength and adaptation exponents on the type of transition to synchrony states. 
\par Therefore, our study has yielded valuable insights into how higher-order interactions, together with adaptivity, influence the emergence of synchronization in multilayer frameworks. However, it is important to recognize that many potential avenues for further investigation remain unexplored. In this context, it will be interesting to study the effect of different adaptation schemes other than order parameter adaptation. The role of inertia in such an adaptive multilayer network with higher-order interactions may also be investigated. Our results could have significant implications for understanding dynamics in brain, biological, and social multilayer networks where both group interactions and adaptations are crucial. In particular, studying how higher-order interactions influence the overall system dynamics could offer a deeper interpretation of these complex networks.

\appendix
\section{Influence of $B$ on the dynamics} 
\label{B_effect}
\begin{figure}[htp] 
    \centering
    \includegraphics[width=\columnwidth]{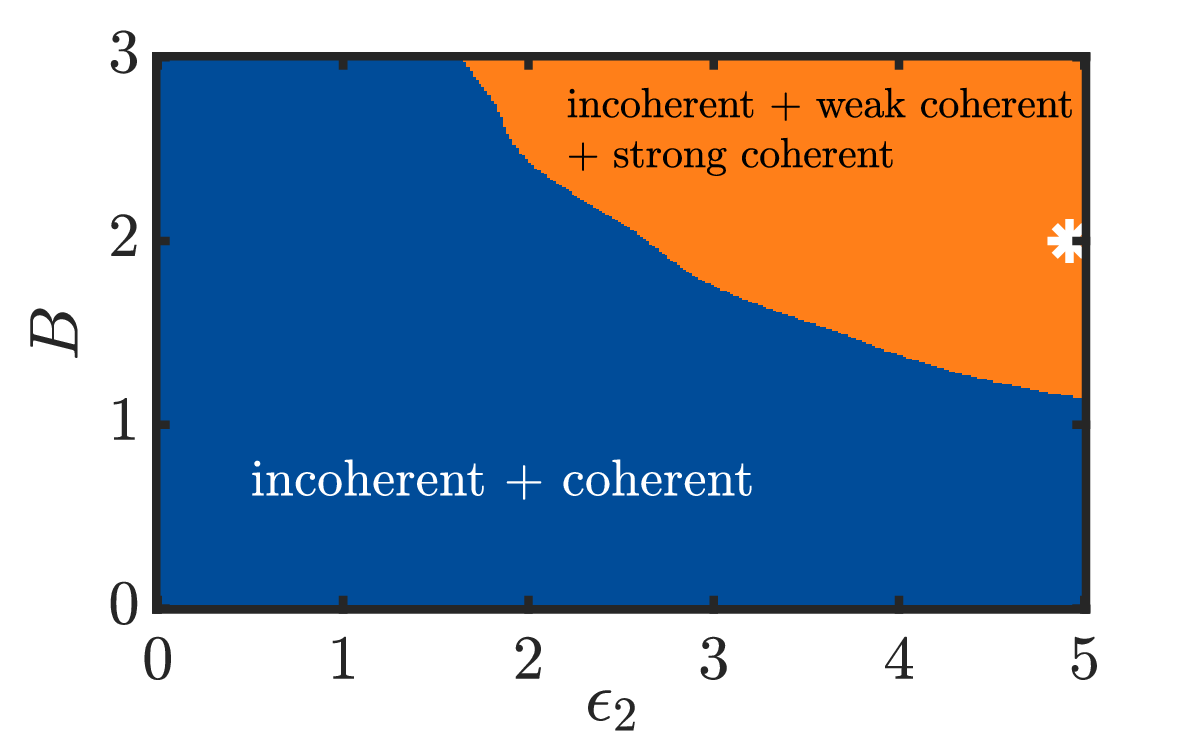}
    \caption{Influence of  constant $B$ and $\epsilon_2$ on the dynamics of the system. The blue regime denotes the bistability of incoherent and coherent states and subsequent explosive transition to synchrony. The orange region shows the multistability of incoherent, weak coherent, and strong coherent states and implies a tiered transition to synchronization. The white star denotes the values at which our calculations for Sec.~\ref{4.1} have been conducted. Hence, it is noted that higher values of $B$ and $\epsilon_2$ promote tiered synchronized states. }
    \label{fig8}
\end{figure}

Here, we illustrate the influence of the constant $B$ of the adaptation function Eq.~\eqref{adaptation_func} on the dynamics of the system with the linear adaptation parameters $p_1=p_2=h_1=h_2=1$ (refer to Section~\ref{4.1}). In Fig.~\ref{fig8}, the interplay of the interactions of the $2$ siplex $\epsilon_2$ and $B$ reveals that weak values of $B$ promote explosive transitions (painted blue) to synchrony due to the increase of bistability of incoherent and coherent states in the system. However, higher values of B ( $B>1$) introduce a multistability (orange region) of incoherent, weak coherent, and strong coherent states for sufficiently strong HOI coupling. Thus, a tiered transition to synchronization is invoked for higher values of $B$ and $\epsilon_2$. The regions associated with different transitions have been distinguished by counting the number of stable solutions at each pair $(\epsilon_{2}, B)$ for varying pairwise coupling $\epsilon_{1}$. If the number of stable solutions is $2$, it denotes a bistability, and hence an explosive transition to synchronization emerges, and if it is more than $3$, it denotes multistability and implies a tiered transition to synchronization. 
\section{Basin of attraction for nonlinear adaptation case} \label{basin_nonlinear}
\begin{figure}[htp]
    \centering
    \includegraphics[width=0.8\columnwidth]{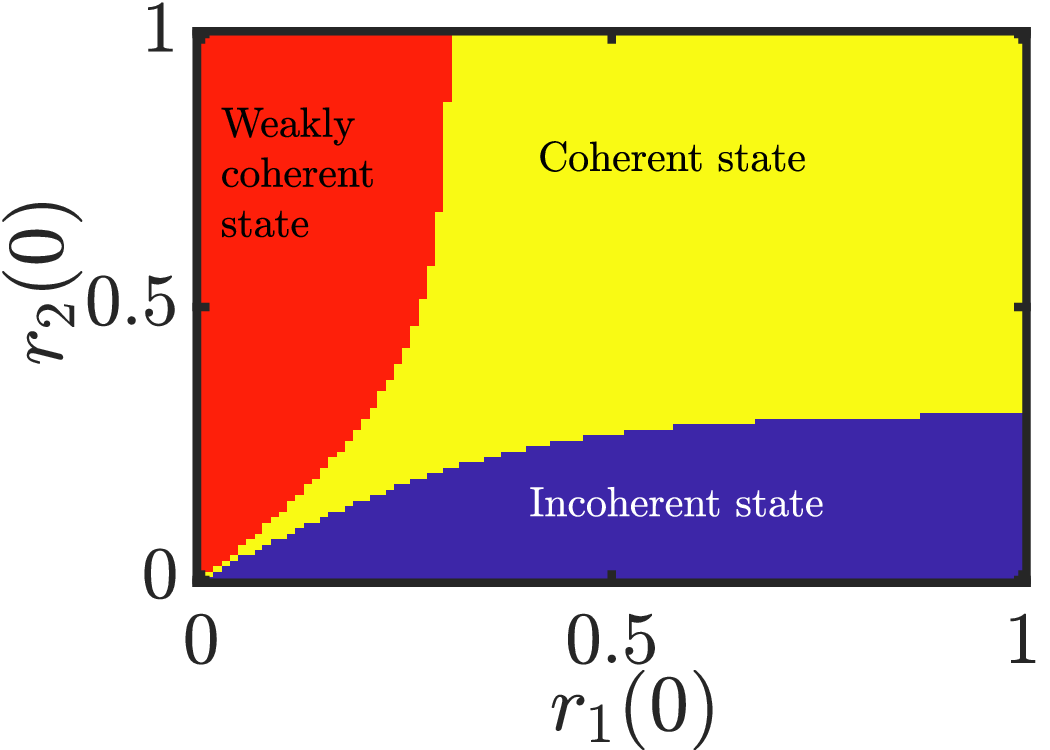}
    \caption{Basin of attraction for the multiple stable states for nonlinear adaptation with parameters $p_1=p_2=-1$,$h_1=h_2=2.0$, $\epsilon_1=6$, $\epsilon_2=0.4$. The purple regime shows the initial conditions for the stable incoherent state, the red region denotes the ones required to attain the weakly coherent state, and the yellow region depicts the initials for the strongly synchronized state. }
    \label{fig9}
\end{figure}

The basin of attraction has been simulated for the case of nonlinear adaptation functions in Fig.~\ref{fig9}.  As noted in Sec~\ref{4.2}, the system exhibits multi-stability of a stable incoherent, weakly coherent, and strongly coherent states at the parameters $p_1=p_2=-1$,$h_1=h_2=2.0$, $\epsilon_1=6$, $\epsilon_2=0.4$. $r_1(0)$ and $r_2(0)$ denote the initial conditions for $r_1$ and $r_2$ since the reduced dimension equations for $r_1$ and $r_2$ (Eq.~\eqref{op_evo_2layer}) are simulated. To achieve the weak synchronized branch, the initial conditions depicted by the red region must be used, i.e., the initials of one layer can be set in strong synchrony while that of the other branch is set in desynchronization. The yellow region depicts the initials required for the appearance of the strong coherent state, while the purple region depicts that for the incoherent one. 

\bibliographystyle{apsrev4-2}
\bibliography{main.bib}  
\end{document}